\documentclass[prb,twocolumn,superscriptaddress,showpacs,preprintnumbers,amsmath,amssymb,floats]
{revtex4}

\usepackage{txfonts}
\usepackage{amssymb}
\usepackage{graphicx}

\begin{document}

\title{ Origin of the kink in the band dispersion of the ferromagnetic perovskite SrRuO$_{3}$: Electron-phonon coupling }

\author{H. F. Yang}
\author{Z. T. Liu}
\author{C. C. Fan}

\affiliation{State Key Laboratory of Functional Materials for Informatics,
Shanghai Institute of Microsystem and Information Technology (SIMIT),
Chinese Academy of Sciences, Shanghai 200050, China}
\affiliation{CAS-Shanghai Science Research Center, Shanghai 201203, China}

\author{Q. Yao}
\affiliation{State Key Laboratory of Functional Materials for Informatics,
Shanghai Institute of Microsystem and Information Technology (SIMIT),
Chinese Academy of Sciences, Shanghai 200050, China}
\affiliation{State Key Laboratory of Surface Physics, Department of
Physics,  and Advanced Materials Laboratory, Fudan University,
Shanghai 200433, China}

\author{P. Xiang}
\author{K. L. Zhang}
\author{M. Y. Li}
\author{H. Li}
\author{J. S. Liu}
\author{D. W. Shen}\email{dwshen@mail.sim.ac.cn}
\affiliation{State Key Laboratory of Functional Materials for Informatics,
Shanghai Institute of Microsystem and Information Technology (SIMIT),
Chinese Academy of Sciences, Shanghai 200050, China}
\affiliation{CAS-Shanghai Science Research Center, Shanghai 201203, China}

\author{M. H. Jiang}
\affiliation{State Key Laboratory of Functional Materials for Informatics,
Shanghai Institute of Microsystem and Information Technology (SIMIT),
Chinese Academy of Sciences, Shanghai 200050, China}

\date{\today}

\begin{abstract}

Perovskite SrRuO$_{3}$, a prototypical conductive ferromagnetic oxide, exhibits a kink in its band dispersion signalling the unusual electron dynamics therein. However, the origin of this kink remains elusive. By taking advantage of the combo of reactive molecular beam epitaxy and \textit{in situ} angle-resolved photoemission spectroscopy, we systematically studied the evolution of the low-energy electronic structure of SrRuO$_{3}$ films with thickness thinning down to nearly two-dimensional limit in a well-controlled way. The kink structure persists even in the 4-unit-cell-thick film. Moreover, through quantitative self-energy analysis, we observed the negligible thickness dependence of the binding energy of the kink, which is in sharp contrast to the downward trend of the Curie temperature with reducing the film thickness. Together with previously reported transport and Raman studies, this finding suggests that the kink of perovskite SrRuO$_{3}$ should originate from the electron-phonon coupling rather than magnetic collective modes, and the in-plane phonons may play a dominant role. Considering such a kink structure of SrRuO$_{3}$ is similar to these of  many other correlated oxides, we suggest the possible ubiquity of the coupling of electrons to oxygen-related phonons in correlated oxides.

\end{abstract}


\maketitle


In correlated oxides, the delicate interplay of charge, spin, lattice and orbital manifested by various collective excitations, gives rise to a wealth of fascinating quantum phenomena, such as metal-insulator transition\cite{RMP_MIT}, high-$T_c$ superconductivity\cite{RMP_Cuprate1,RMP_Cuprate2}, colossal magnetoresistance\cite{RMP_CM}, and multiferroics\cite{Multiferroics}. Among these compounds, perovskite SrRuO$_{3}$ --- a moderately correlated conductive ferromagnet\cite{Singh1997, RMP_SRO, 1958} --- has attained continuous interests. On the application side, it is widely utilized as conductive electrodes due to the good stability and structurally compatibility with other correlated oxides\cite{RMP_SRO}; meanwhile, it is explored to be a key integrant in fabricating oxide heterostructures/superlattices\cite{SL_1, Darrel, SL_2, SL_3, SL_4, SL_5, SL_6, SL_7, SL_8}, which may contribute to new functionalities in electronics and spintronics\cite{RMP_SRO}. From the viewpoint of fundamental studies, SrRuO$_{3}$ is a simple but profound model system to explore how many-body interactions determine the physical properties\cite{KM_Shen, limit, Hund, Optical, J_Millis}, and the underlying mechanism may provide a hint on novel physics of other correlated oxides including the unconventional superconductivity in Sr$_{2}$RuO$_{4}$\cite{RMP_214} and quantum criticality in Sr$_{3}$Ru$_{2}$O$_{7}$\cite{327}.

For correlated oxides, a kink in the low-energy band dispersion reflects the unusual electron dynamics --- the scattering rate of electrons has been altered within a narrow energy range --- which usually implies the existence of pronounced coupling between electrons and various collective excitations\cite{RMP_Cuprate2, 214 327, SVO, LSMO, Cuprate, Zhou XJ}. Thus, the interpretation of the kink is of significance to understand the essential effects on physical properties of correlated systems. For example, there have been lasting and intense debates on whether the kink-like feature discovered in the dispersion of cuprate superconductors stems from spin fluctuations or lattice waves (phonons), which is intimately related with the fundamental understanding of the pairing mechanism of high-$T_c$ superconductivity\cite{Cuprate, Zhou XJ, RMP_Cuprate2}. For SrRuO$_{3}$, previous studies have reported a kink-like feature in its band dispersion\cite{KM_Shen}, which suggests remarkable renormalization of low-energy electrons by some collective modes. However, the details of which mode is most relevant, has not been completely explored yet till now.

In this article, we took advantage of the reactive molecular beam epitaxy (MBE) to fabricate a series of high-quality SrRuO$_{3}$ thin films with well-controlled thickness. Through tuning the extra knob of dimensionality, we can effectively adjust the ferromagnetism and meanwhile keep the basic crystal structure of films, which is expected to result in the pronounced dichroism between the possible electron-phonon and electron-magnon interactions in films. Consequently, we performed a systematic \textit{in situ} angle-resolved photoemission spectroscopy (ARPES) study of the kink structure to pin down which boson modes are most relevant. We found that the kink persists in the film as thin as 4 unit cells (uc). Through quantitative self-energy analysis, we found that, while the film thickness was progressively reduced, the kink energy remained to be around 62 meV below the Fermi level ($E_F$). This observation, together with reported thickness dependent Curie temperature ($T_{C}$) and Raman studies of phonons, suggests that the kink in SrRuO$_{3}$ originates from the the strong electron-phonon coupling, and the collective magnetic excitations may have little contribution. By referring other correlated oxides exhibiting similar kink-like structure, we infer that the electron-phonon coupling may be ubiquitous in correlated oxides.


\section{Experimental}

\begin{figure}
\includegraphics[width=8.5cm]{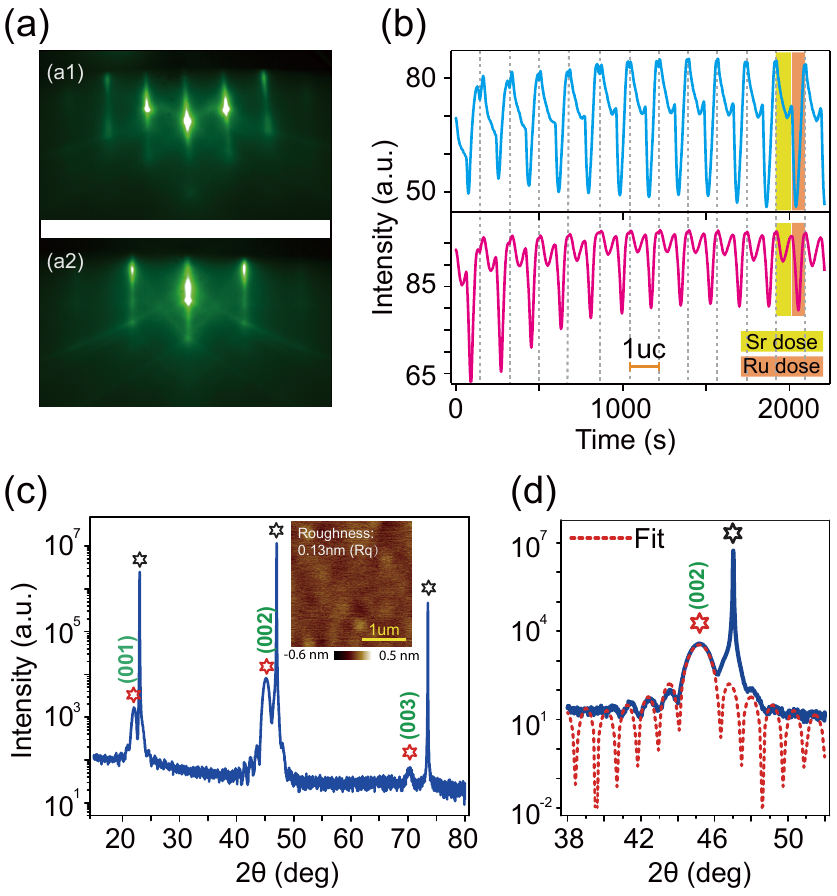}
\caption{(Color online) MBE growth and characterization of SrRuO$_{3}$ thin films with well-controlled thicknesses. (a) Typical RHEED patterns of a 21-uc-thick SrRuO$_{3}$ epitaxial thin film. (a1) and (a2) are taken along [100]$_{p}$ and [110]$_{p}$ azimuth directions, respectively. (b) RHEED intensity curve: time dependence of the intensities of [00] and [01] diffraction spots in the [100]$_{p}$ RHEED pattern (a1). Each oscillation framed by two adjacent dotted lines represents the growth of 1 uc SrRuO$_{3}$ thin film. (c) The typical X-ray diffraction $\theta$-2$\theta$ scan of a 21 uc SrRuO$_{3}$ thin film whose nominal thickness is determined by counting the number of RHEED oscillations. Red markers represent diffraction peaks of the film, while black ones represent those of the NdGaO$_3$ substrate. The inset shows the morphology of a typical 4 uc SrRuO$_{3}$ thin film measured by atomic force microscope, which gives a rather flat surface with a roughness of only 0.13 nm (Rq). (d) A close-up of the 002 diffraction peak of this 21 uc SrRuO$_3$ film, showing clear Keissig fringes. The red dotted line is the best fit to the data, which gives the thickness of 21 uc, in good agreement with the nominal thickness determined by counting the RHEED oscillations.}
\end{figure}

High-quality single crystalline (001)$_{p}$ (where $p$ denotes pseudocubic indices) SrRuO$_{3}$ thin films with a spectrum of thicknesses were fabricated on (001)$_{p}$ NdGaO$_{3}$ substrates (without special treatment; mixed termination) using a reactive MBE system (DCA R450). The in-plane lattice constant of (001)$_{p}$ NdGaO$_{3}$ is 3.865 {\AA}, which leads to a 1.65\% compressive strain to the pseudocubic SrRuO$_{3}$ (3.93 {\AA}). The shuttered growth mode was applied to synthesize films in the distilled ozone atmosphere of 1.0 $\times$ 10$^{-6}$ Torr. During the growth, the temperature of substrates was kept at 600 $^\circ$C according to the thermocouple behind the sample stage. Moreover, the overall growth rate and surface structure of thin films were monitored by \emph{in situ} reflection high-energy electron diffraction (RHEED) during growth. Strontium (Sr) and ruthenium (Ru) were evaporated from a thermal Knudsen cell and an electron-beam evaporator with delicate feedback-control of flux, respectively. The Sr and Ru fluxes were approximately 1.2$\times$10$^{13}$ atoms$/$(cm$^{2}$s), which were checked before and after the deposition by a quartz crystal microbalance.

Fig. 1(a1) and (a2) display the typical RHEED patterns of a 21-uc-thick SrRuO$_{3}$ epitaxial thin film taken along [100]$_{p}$ and [110]$_{p}$ azimuth directions, respectively. The remarkable shiny spots and visible Kikuchi lines reveal the fine surface with long-range ordered lattice structure. The amplitudes and periods of RHEED intensity oscillations of the [00] and [01] diffraction rods tend to be constant, indicating the layer-by-layer and stoichiometric growth of this film\cite{OMBE}. Note that each oscillation represents the growth of one layer of SrRuO$_{3}$, as depicted in Fig. 1(b). Consequently, closing/opening the Sr/Ru shutters according to the RHEED intensity curve can intentionally control the thickness of thin film. In this way, the thicknesses of all SrRuO$_{3}$ films in this work were well controlled. As shown in the inset of Fig. 1(c), the morphology of a typical 4-uc-thick film measured by atomic force microscope illustrates a surface roughness of only 0.13 nm, which guarantees the well-controlled thickness of the ultrathin films and following ARPES measurements. The typical X-ray diffraction (XRD) of the SrRuO$_{3}$ film is shown in Fig. 1(c). The $\theta$-2$\theta$ scan result is consistent with the growth of phase-pure (001)$_{p}$-oriented SrRuO$_{3}$. The clear Kiessig interface fringes indicate the sharp and smooth interfaces between the thin film and substrate. The scheme of the thickness control could be further confirmed by the explicit fitting of Keissig fringes in the XRD pattern\cite{Laue fit}. For example, the best fit to the $\theta$-2$\theta$ scan around the 002 diffraction peak [Fig. 1(d)] gives a thickness of 21 uc, which is in good agreement with that determined by counting RHEED oscillations.

After growth, the film was cooled down to 200 $^{\circ}$C and then immediately transferred to the ARPES chamber within 5 minutes through a transfer chamber bridging the MBE and ARPES with a vacuum of around 1.0 $\times$ $10^{-10}$ Torr. In so doing, clean sample surface is expected to be preserved\cite{KM_Shen}. Then ARPES measurements were performed with a Specs UVLS monochromatized helium discharging lamp (He I$\alpha$, 21.2 eV) and a VG Scienta R8000 analyser under an ultra-high vacuum of 8.0 $\times$ $10^{-11}$ Torr. Thin films were measured at 14 K with an energy resolution of 10.5 meV and angle resolution of 0.3$^{\circ}$. The $E_F$ was referenced to that of a polycrystalline gold newly deposited and electrically connected to samples. We note that finite density of states near $E_F$ persists in even 2-uc-thick film, which is in line with the best SrRuO$_{3}$ films ever reported\cite{limit}. In order to avoid the photoemission charging effect due to the insulating behavior of the NdGaO$_{3}$ substrate, conducting silver paste was used to carefully cover edges of the substrate. In all cases, we experimentally conformed that no charging effect occurred through photon flux test. Also, no apparent signs of degradation were observed during measurements.

\section{Results and Discussions}

\begin{figure}[t]
\includegraphics[width=8.5cm]{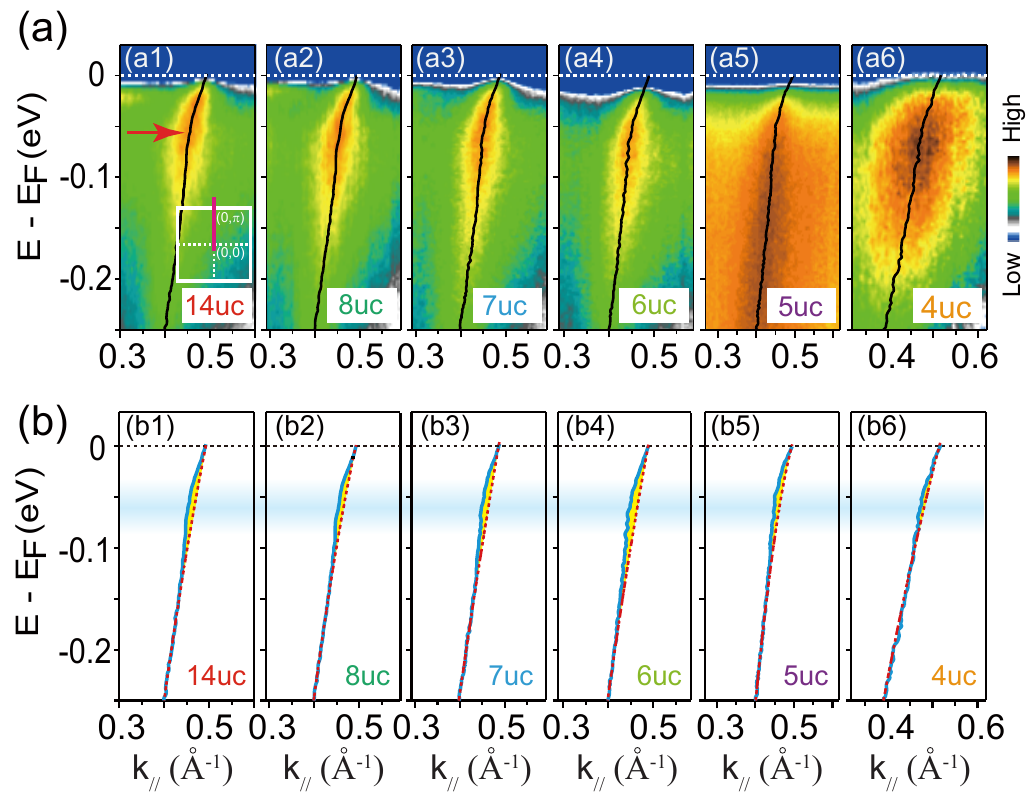}
\caption{(Color online) The kink exists in all SrRuO$_{3}$ films down to 4-uc-thick. (a) The photoemission intensity spectra of SrRuO$_{3}$ films with various thicknesses near $E_F$ along the (0, 0)-(0, $\pi$) high-symmetry direction as shown by the inset of a1. The black lines overlaid are extracted band dispersions obtained through fitting MDCs spaced by 3 meV with Lorentz functions. Red arrow in a1 roughly marks the energy position of the kink. (b) Extracted band dispersions with bare bands obtained by polynomial fittings with an order of two as indicated by dotted red lines. The difference between the extracted band dispersion and the polynomial bare band is treated as the self energy denoted by the yellow shadow. The kink structure is present in all films ranging from 14 to 4 uc. }
\end{figure}

\begin{figure}[t]
\includegraphics[width=8.5cm]{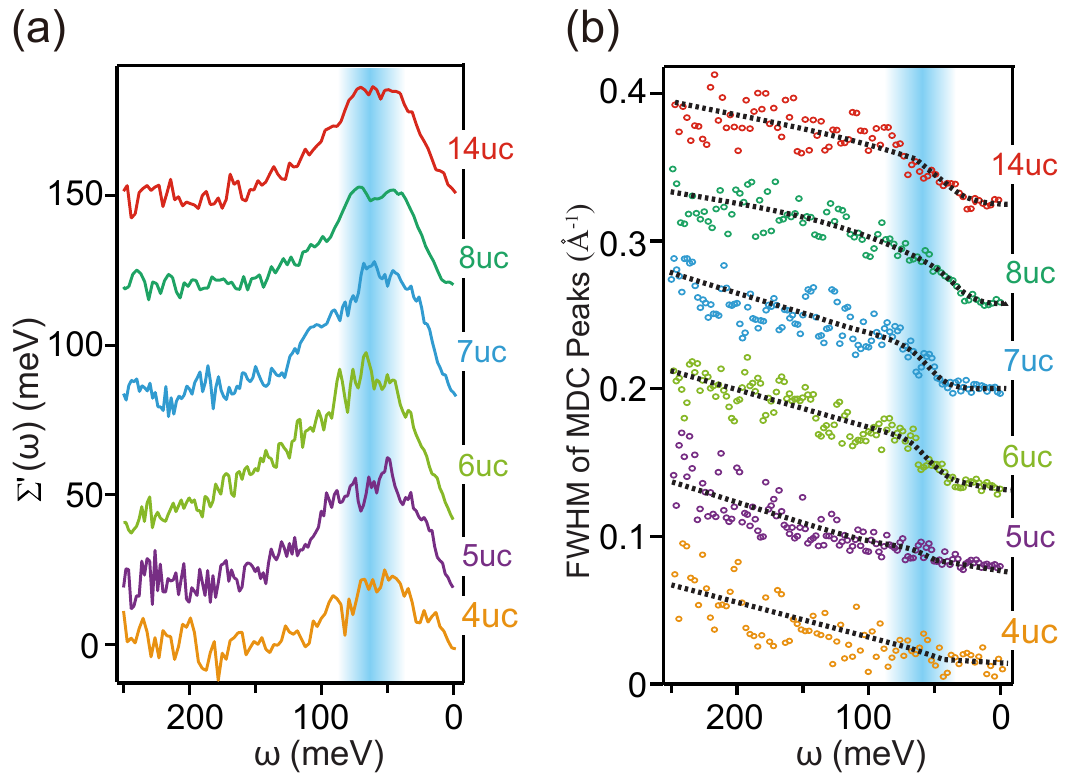}
\caption{ (Color online) Negligible thickness dependence of the kink in SrRuO$_{3}$. (a) Real part ($\Sigma$$^{\prime}$($\omega$)) analyses of the self energies of films. Offsets are used to display those curves clearly, and no normalizations are applied. The gradient cyan shadow area indicates the maxima region of the real part of the self energies, which denotes the binding energies of the kinks. All kinks occur at around 62 meV below the E$_{F}$. (b) Corresponding imaginary part ($\Sigma$$^{\prime\prime}$($\omega$), can be represented by the FWHM of MDC peaks) analyses of self energies. To display these data clearly, offsets and normalizations of ``amplitude'' (difference between the biggest and smallest FWHMs in one film) are used. Besides, black dotted lines are used to be guides to the eyes. The gradient cyan shadow marks the abrupt change of the imaginary part, which represents the kink. All kinks occur at around 62 meV below the E$_{F}$, which is in accordance with the real part analyses in a. }
\end{figure}

Fig. 2(a1) displays the band dispersion of one typical 14-uc-thick  SrRuO$_{3}$ film along the (0,0)-(0,$\pi$) high symmetry direction (see the inset). Consistent with the previous report\cite{KM_Shen}, the quasi-1D band $\beta$ exhibits sharp quasi-particle peaks, indicating that the sample is indeed in the Fermi liquid regime. The overlaid black curve is the exacted dispersion by fitting momentum distribution curves (MDCs) with Lorentzian functions. It is evident that the slope of the dispersion near $E_F$ is significantly smaller than that of high-binding-energy part (-0.15 eV to -0.25 eV). Such an abrupt change of the slope is the key characteristic of the kink structure signalling significant electron-boson coupling\cite{KM_Shen, EP_ZX, NPhys}. As displayed in Figs. 2(a1-a6), while the film thickness keeps decreasing, the kink structure persists in the dispersion of the band $\beta$, and can even be observed in a 4-uc-thick film [Fig. 2(a6)], aside from the broader background and smaller slope of high-binding energy part partially due to the increasing electron correlations induced by reducing the dimensionality\cite{2D limit}. For films with less than 4 uc, the dispersion is too broad to extract the kink structure, while finite spectral weight near the $E_F$ still remains in 2-uc-thick films (not shown).

To identify such kink evolution with dimensionality in an accurate way, quantitative self-energy analysis was carried out\cite{KM_Shen, SVO}. Here the self-energy is handled in a practical way, which was proved to be effective in identifying the kink\cite{KM_Shen, SVO}. We first fitted the band $\beta$ (specifically the high binding-energy part) with a polynomial function with an order of two to approximately yield the bare band dispersion marked by red dotted lines in Figs. 2(b1-b6); consequently, by subtracting the exacted band with the fitted bare band, we can obtain the the real part of the self-energy $\Sigma$$^{\prime}$($\omega$), as marked by yellow shadows in Fig. 2(b). Fig. 3(a) illustrates the thickness dependence of $\Sigma$$^{\prime}$($\omega$). Each curve evidently shows a maxima which can be used to mark the energy position of the kink. It is found that the kink structure is always located at around 62 meV below $E_F$, as highlighted by the gradient cyan shadow. Such a result can be also confirmed by the imaginary part of the self-energy $\Sigma$$^{\prime\prime}$($\omega$) analysis, which is directly related with momentum distribution curve (MDC) peak widths [see Fig. 3(b)]. As we can see, each $\Sigma$$^{\prime\prime}$($\omega$) curve exhibits a characteristic step-like increase, implying the sudden change of the scattering rate of electrons. With the decreasing of the film thickness, the step-like feature's characteristic energy remains around 62 meV (binding energy) as well, in line with the $\Sigma$$^{\prime}$($\omega$) analysis. We can further estimate the kink energy of each film by applying Gaussian fitting to the maxima region of the $\Sigma$$^{\prime}$($\omega$), given that the maxima region of the $\Sigma$$^{\prime}$($\omega$) qualitatively adopts the Gaussian function shape. As shown by the left axis of Fig. 4a, the kink energy qualitatively remains unchanged with film thickness.

Perovskite SrRuO$_{3}$ was reported to host a fundamental thickness limit of the long-range ferromagnetism: the Curie temperature T$_{C}$ keeps decreasing with thinning the film \cite{limit}, as illustrated by the reproduced data in Fig. 4(a) (the right axis). In a typical itinerant ferromagnet, the effective exchange interaction $J$ is expected to be proportional to the $T_C$, and thus the energy of spin waves or magnons, which is proportional to the spin wave stiffness $D$ $\propto$ $J$, would be reduced as the $T_C$ decreasing. If the kink structure of SrRuO$_{3}$ indeed originates from the electron-magnon coupling, the energy of the kink should have shown an evident downward trend with thinning the film. However, by contrast, the binding energy of the kink shows negligible thickness dependence as illustrated in Fig. 4(a) (the left axis). In addition, for another two members of Ruddlesden-Popper (RP) series of ruthenates, Sr$_2$RuO$_4$ and Sr$_3$Ru$_2$O$_7$, whose the magnetic ground states are totally different from SrRuO$_{3}$, the spectral behaviors of the kinks are similar to that of SrRuO$_{3}$\cite{214 327}. Therefore, we suggest that the collective magnetic excitations may have little contribution to the kink of SrRuO$_{3}$. Alternately, it was proposed that strong electron-electron interaction could solely lead to the formation of a kink in the electron band\cite{NPhys, E-E} as well, which however is not the case of SrRuO$_{3}$ since SrRuO$_{3}$ is known as a moderately correlated without well-separated sub-Hubbard bands\cite{RMP_SRO}.

\begin{figure}[t]
\includegraphics[width=8.5cm]{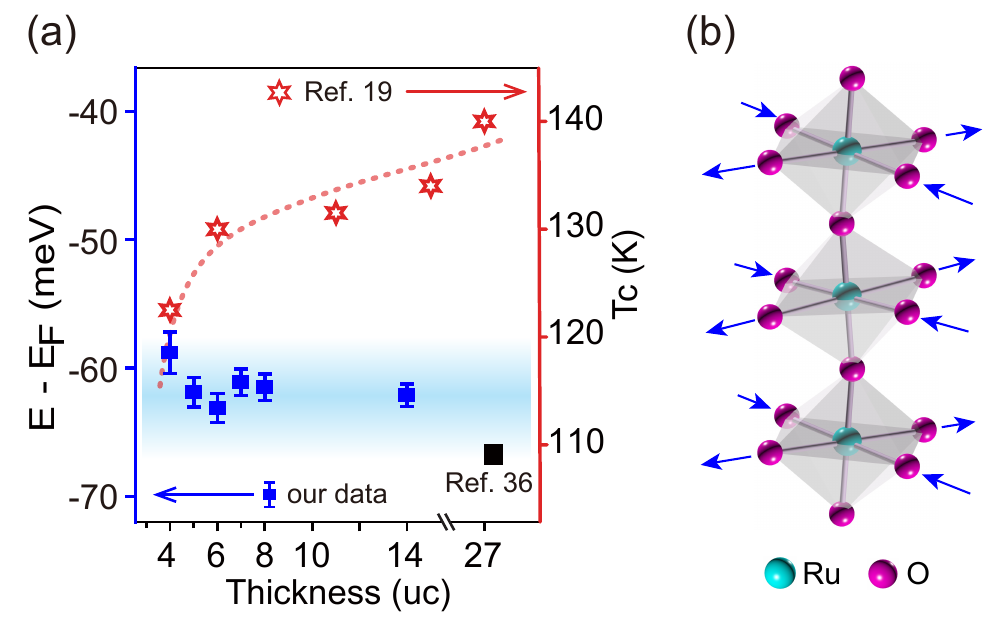}
\caption{ (Color online) Electron-phonon coupling origin of the kink in SrRuO$_{3}$. (a) Left axis: measured negligible thickness dependence of the energy position of the kink. The kink energy is extracted by applying Gaussian function fitting to maxima region of the real part of self energy. Right axis: reported thickness dependence of the Curie temperature (ref. 19). The black rectangular marker displays the energy scale of in-phase stretching phonon mode proposed by the combined studies of Raman spectroscopy and calculation (ref. 36). (b) Schematic of in-phase stretching phonon mode (associated with the movement of in-plane oxygen atoms as marked by blue arrows) (ref. 36), based on the perovskite structure of SrRuO$_{3}$ which is constituted of oxygen octahedra connected by oxygen atoms (note that Sr atoms are not drawn for simplicity). }
\end{figure}

On the other hand, the combined study of Raman spectroscopy and lattice dynamical calculations revealed that perovskite SrRuO$_{3}$ films possess the prominent in-phase stretching mode in the phonon spectra [as illustrated in Fig. 4(b) schematically], the energy scale of which (540cm$^{-1}\sim$ 67 meV) is remarkably close to our measured kink's binding energy 62 meV\cite{Raman}. Besides, such bonding-stretching phonons with similar energy scale have been well identified as the origin of the kink structure in the band dispersion of RP-structured manganite La$_{2-2x}$Sr$_{1+2x}$Mn$_{2}$O$_{7}$ \cite{LSMO}. Thus, it is most probable that the coupling of electrons to the bond-stretching phonons plays a dominant role in the formation of kink in perovskite SrRuO$_{3}$ films. Indeed, since in-plane phonon mainly involves in-plane movements of atoms, its mode energy should not notably change with driving SrRuO$_{3}$ to two-dimensional limit.

As mentioned above, for two other members of RP-structured ruthenates, Sr$_{2}$RuO$_{4}$ and Sr$_{3}$Ru$_{2}$O$_{7}$ both show the kink structure with the same binding energy in the band dispersion\cite{214 327}. Considering their more two-dimensional electronic structure and negligible thickness dependence of kink in perovskite SrRuO$_{3}$ think films, it is natural to assign these phenomena to the coupling of electrons to the \emph{in-plane} stretching phonon branch. Moreover, such kink structure in the band dispersion was also observed in many other correlated oxides, such as various cuprate superconductors\cite{Cuprate, Zhou XJ} and SrVO$_{3}$\cite{SVO}. In terms of crystal structure, these compounds belong to perovskite and layered perovskite oxide family, which are constituted of oxygen octahedra/tetrahedra connected by oxygen atoms [see the structure of SrRuO$_{3}$ in Fig. 4 (b) as an example]. The vibration of oxygen atoms, together with various tilts and distortions of similar oxygen octahedra would lead to considerable phonon modes. Once such phonon modes couple with electrons, renormalization with cut-off binding energies may occur and scattering rate near the $E_F$ may be changed as well, thus forming a kink in the band dispersion.

Thus, we suggest that the interaction between electrons and phonons may be ubiquitous (in many cases even form a kink in the band dispersion, as discussed above) and then impact the physical performance of correlated oxides with the transition metal-oxygen octahedra units. Particularly, this kind of coupling has also been considered to be important in understanding the the formation of Cooper pairs in cuprates\cite{EP_ZX, Cuprate}, and the colossal magnetoresistance effect in manganites\cite{LSMO}. The studies of electron-phonon coupling in ruthenates may provide opportunities to uncover the roles phonons play in these novel materials.


\section{Conclusion}

To summarize, we have utilized the delicate combo of reactive MBE and \textit{in situ} ARPES to systematically study the kink structure of SrRuO$_{3}$ thin films with various well-controlled thicknesses. The kink structure exists in all SrRuO$_{3}$ films (even in 4-uc-thick films). Moreover, through quantitative self-energy analysis, we observed that all kinks approximately occur at around 62 meV below the E$_{F}$. This observation, and comparisons with other oxides, together with reported transport and Raman studies, demonstrate that the kink originates from the the strong electron-phonon coupling other than electron-magnon coupling. In addition, we discussed the possible ubiquity of electron-phonon coupling in correlated oxides. Our work not only uncovers the kink origin of SrRuO$_{3}$ (for the first time) and gives a hint on fundamental exploration of other correlated oxides, but also demonstrates that the \textit{in situ} MBE and ARPES system could be a powerful toolkit in studying novel quantum materials.

\section{ACKNOWLEDGMENTS}

We gratefully acknowledge the helpful discussions with Prof. Xiangang Wan. This work was supported by National Basic Research Program of China (973 Program) under the grant No. 2012CB927401 and the National Science Foundation of China under Grant Nos. 11274332 and 11227902. H. F. Yang and D. W. Shen are also supported by the ``Strategic Priority Research Program (B)'' of the Chinese Academy of Sciences (Grant No. XDB04040300).

\end{document}